\documentclass[a4paper,12pt]{article}
\usepackage[latin2]{inputenc}
\usepackage[T1]{fontenc}
\usepackage{times}

\begin{document}

\title{Horizon conformal entropy in\\ Gauss-Bonnet gravity}

\author{M. Cvitan\thanks{E-mail: mcvitan@phy.hr},
  S. Pallua\thanks{E-mail: pallua@phy.hr} and
  P. Prester\thanks{E-mail: pprester@phy.hr}\\[5mm]
  \normalsize \it Department of Theoretical Physics,\\[-1mm]
  \normalsize \it Faculty of Natural Sciences and Mathematics,\\[-1mm]
  \normalsize \it University of Zagreb,\\[-1mm]
  \normalsize \it Bijenička c. 32, pp. 331, 10000 Zagreb, Croatia}
\date{\normalsize Preprint ZTF 02-03}

\maketitle

\begin{abstract}
We treat spherically symmetric black holes in Gauss--Bonnet gravity by
imposing boundary conditions on fluctuating metric on the horizon.
Obtained effective two-dimensional theory admits Virasoro algebra near
the horizon. This enables, with the help of Cardy formula, evaluation
of the number of states. Obtained results coincide with the known
macroscopic expression for the entropy of black holes in Gauss--Bonnet
gravity.
\end{abstract}

\section{Introduction}

The well-known Bekenstein--Hawking (BH) formula \cite{BekHaw} connects
area of the black hole horizon with its entropy, i.e.,
\begin{equation}\label{sbeha}
S_{\mbox{\scriptsize BH}}=\frac{A}{4\hbar G}\;.
\end{equation}
A considerable research effort in recent years was performed in order
to understand microscopic interpretation of this relation. A
particularly promising approach seems to be based on conformal field
theory and Virasoro algebra. In fact, it was realized by Brown and
Henneaux \cite{BroHen86} that in $2+1$ dimensions and after imposing
asymptotic AdS3 symmetry one can identify two copies of Virasoro
algebra and corresponding central charges. Further analysis
\cite{BanStro} has reproduced Bekenstein--Hawking entropy for black
holes in this theory. More recently, several papers addressed the
problem of $D$-dimensional black holes. In particular, Solodukhin is
treating \cite{Solod99} the spherically symmetric black holes with the
metric
\begin{equation}\label{gsph}
ds^2=\gamma_{ab}(x)dx^adx^b+r(x)^2d\Omega_{D-2}\;,
\end{equation}
where $d\Omega_{D-2}$ is metric on $(D-2)$-dimensional sphere of unit
radius. In this approach one considers fluctuations of the field
$r(x)$ on a two-dimensional space-time with the metric
$\gamma_{ab}(x)$. The author was able to identify a particular group
of diffeomorphisms under which the horizon is invariant. The Einstein
action reduces to a two-dimensional action of Liouville type. One is
able to identify a Virasoro algebra. The aim is then to calculate the
entropy from Cardy formula \cite{Cardy}
\begin{equation} \label{carfo}
S_{\mbox{\scriptsize C}}=2\pi\sqrt{\left(\frac{c}{6}-
4\Delta_{g}\right)\left(\Delta-\frac{c}{24}\right)}\;,
\end{equation}
where $\Delta$ is the eigenvalue of Virasoro generator $L_0$ for the
state we calculate the entropy and $\Delta_g$ is the smallest
eigenvalue. It was shown that the corresponding entropy reproduces BH
result (\ref{sbeha}). Another approach is due to Carlip
\cite{Carlip98,Carlip99,Carlip02,Park01,Silva02} where one requires
in $D$-dimensional gravity a set of boundary conditions near horizon.
That leads to central extension for the constraint algebra of general
relativity. Due to assumed boundary conditions this algebra contains
Virasoro algebra whose existence enables one to
calculate conformal charge and via Cardy formula (\ref{carfo}) the
entropy. All these papers confirm that microscopic description via
conformal theory reproduces the classical BH result for Einstein
gravity. The question which we want to investigate in this Letter is
if such description reproduces the classical result also for theories
which differ from Einstein action by new terms written in terms of
products of Riemann tensors and corresponding covariant derivatives.
In fact it is known that the classical entropy differs from the BH
formula in these cases. Introduce e.g., the (extended) Gauss--Bonnet
(GB) densities
\begin{equation}\label{lgbm}
\mathcal{L}_m(g)=\frac{(-1)^m}{2^{m}}\delta_{\mu_1\nu_1\ldots
\mu_m\nu_m}^{\rho_1\sigma_1\ldots\rho_m\sigma_m}
{R^{\mu_1\nu_1}}_{\rho_1\sigma_1}\cdots
{R^{\mu_m\nu_m}}_{\rho_m\sigma_m}\;,
\end{equation}
where $R_{\mu\nu\rho\sigma}$ is Riemann tensor for metric
$g_{\mu\nu}$ and $\delta_{\alpha_1\ldots\alpha_k}^{\beta_1\ldots
\beta_k}$ is totally antisymmetric product of $k$ Kronecker deltas,
normalized to take values 0 and $\pm 1$.
By definition, we take $\mathcal{L}_0=1$ (cosmological constant term).
Notice also that $\mathcal{L}_1=-R$, i.e. ordinary Einstein action.
General GB action (also known as Lovelock gravity \cite{Lovelock}) is
now given as
\begin{equation}\label{igb}
I_{\mbox{\scriptsize GB}}=
-\sum_{m=0}^{[D/2]}\lambda_m\int d^D x\sqrt{-g}\mathcal{L}_m(g)\;,
\end{equation}
where $g=\det(g_{\mu\nu})$ and $[z]$ denotes integer part of $z$.
Explicit expresion for the entropy of general stationary black hole in
GB theory is \cite{JacMye93}
\begin{equation}\label{sgb}
S_{\mbox{\scriptsize GB}}=\frac{4\pi}{\hbar}\sum_{m=1}^{[D/2]}
m\lambda_m\oint d^{D-2}x\sqrt{\tilde{g}}
\mathcal{L}_{m-1}(\tilde{g}_{ij})\;,
\end{equation}
where the integration can be made on any ($D-2$)-dimensional spacelike
slice of the Killing horizon and $\tilde{g}_{ij}$ is the induced
metric on it. In fact, classical expression for entropy in any
generally covariant gravity theory have been suggested \cite{genent}.

In this Letter we shall investigate in particular the Gauss--Bonnet
action. This action has in fact many interesting properties:
\begin{itemize}
\item In $D$-dimensional space all terms for which $m>D/2$ are
identically equal to zero, because maximal rank of antisymmetric
tensor in such space is $D$. It follows that there is always finite
number of terms in the GB action (which we already included in the
definition (\ref{igb})). Term $m=D/2$ is a topological term. In
fact it is the original Gauss--Bonnet term which exists in even
dimensional spaces and which (with apropriate surface term added) is
equal to the Euler character of that space. So, only terms for which
$m<D/2$ are contributing to equations of motion. It means that in
$D=4$ GB action is (neglecting topological effects) just the Einstein
action.
\item \emph{Only} GB terms have the property that resulting equations
of motion contain no more than second derivative of metric
\cite{Lovelock}. They are also free of ghosts when expanded not only
about flat space \cite{noghost} but also about some Randall--Sundrum
brane solutions in $D=5$ \cite{ChNeWe02}.
\item It has a good boundary value problem \cite{Myers87}, in the
sense that we can add surface terms such that the action can be
extremized on space $M$ while keeping only the metric fixed on the
boundary $\partial M$ (if non-GB terms are present in the action we
have to also fix derivatives of components of the metric tensor on 
$\partial M$).
\item Analysis of spherically symmetric classical solutions in empty
space is almost as simple as for pure Einstein case. But, unlike the
Einstein case where there was unique solution (Schwarzschild), for
general GB action there are black hole solutions having more
complicated global topologies with multiple horizons and/or naked
singularities \cite{MyeSim88}.
\item The entropy of GB black holes can be written (at
least in stationary cases) as a sum of {\em intrinsic} curvature
invariants integrated over a cross section of the horizon. As far as
is known only GB actions have this property. Interesting property that
the entropy (\ref{sgb}) has the same form as the action (\ref{igb})
can be described as dimensional continuation of the Gauss--Bonnet
theorem.
\item The entropy of GB black holes is negative for some region of
parameter space. It is speculated that this is connected with the
existence of a new type instability \cite{CvNoOd02}.
\item It can be supersymmetrised. 
\item It is nonrenormalisable.
\end{itemize}
This properties suggest that GB action could be considered as a
natural generalisation of Einstein action.

We shall investigate the entropy problem for this action with the
Solodukhin method. We describe first the simpler case with only
quadratic terms in Riemann tensor and consider spherically symmetric
black holes with fixed boundary conditions for the fluctuating metric.
We calculate the corresponding effective two-dimensional theory.
It will be possible to find Virasoro algebra corresponding to the
diffeomorphisms which preserve above boundary condition. Calculations
of central charge and application of Cardy formula will determine
entropy. We shall find that number of states obtained in such a way
reproduces the classical result of Jacobson and Myers. In the Section
3 we generalise these results to the most general GB theory. In the
last section we end with concluding remarks.

\section{Effective CFT near the horizon}

Now we turn our attention to particular microscopic derivation of
``macroscopic'' expression (\ref{sgb}) for entropy of black holes in
GB theory. For simplicity in this section we put $\lambda_m=0$ for
$m>2$. General action will be considered in the next section. We also
take $\lambda_0=0$ (cosmological constant), because we shall see that
this term is irrelevant for our calculation. In this case action
(\ref{igb}) becomes
\begin{equation}\label{igb2}
I_{\mbox{\scriptsize GB}}=\int d^Dy\sqrt{-g}\left[\lambda_1 R
-\lambda_2\left( R_{\mu\nu\rho\sigma}R^{\mu\nu\rho\sigma}
-4R_{\mu\nu}R^{\mu\nu}+R^2\right)\right]\;.
\end{equation}
Coupling constant $\lambda_1$ is related to more familiar
$D$-dimensional Newton gravitational constant $G_D$ through
$\lambda_1=(16\pi G_D)^{-1}$.

Following Solodukhin \cite{Solod99} we neglect matter and consider
$S$-wave sector of the theory, i.e., we consider only radial
fluctuations of the metric. It is easy to show that in this case
(\ref{igb2}) can be written in the form of an effective
two-dimensional ``higher-order Liouville theory'' given with
\begin{eqnarray}
I_{\mbox{\scriptsize GB}}&=&(D-2)(D-3)\Omega_{D-2}
\int d^2x\sqrt{-\gamma} \nonumber \\
&&\times\Bigg\{2\lambda_2(D-4)r^{D-5}(\nabla r)^2\nabla^2r 
+\lambda_2(D-4)(D-5)r^{D-6}(\nabla r)^4 
\nonumber \\ && \quad
-\left[\lambda_1r^{D-4}+2\lambda_2(D-4)(D-5)r^{D-6}\right](\nabla r)^2
\nonumber \\ && \quad +\left[\frac{\lambda_1r^{D-2}}{(D-2)(D-3)}+
2\lambda_2r^{D-4}\right]\mathcal{R} \nonumber \\ && \quad 
-\left[\lambda_1r^{D-4}+\lambda_2(D-4)(D-5)r^{D-6}\right]
\Bigg\}\;. \label{gb2ef}
\end{eqnarray}
We now suppose that black hole with horizon \emph{is existing} and we
are interested in fluctuations (or better quantum states) near it. In
the spherical geometry apparent horizon $\mathcal{H}$ (a line in
$x$-plane) can be defined by the condition \cite{Russo95}
\begin{equation}\label{hoco}
\left.(\nabla r)^2\right|_\mathcal{H}\equiv
\left.\gamma^{ab}\partial_a r\partial_b r\right|_\mathcal{H}=0 \;.
\end{equation}
Notice that (\ref{hoco}) is invariant under (regular) conformal 
rescalings of the effective two-dimensional metric $\gamma_{ab}$. Near
the horizon (\ref{hoco}) is approximately satisfied.
It is easy to see that near the horizon first two terms in
(\ref{gb2ef}) are suppressed by a factor $(\nabla r)^2$ relative to
the third term (to see this just partially integrate latter and 
discard surface terms) and may be neglected.

If we make reparametrizations
\begin{equation}\label{rfi}
\phi\equiv\frac{2\Phi^2}{q\Phi_h} \;, \qquad
\tilde{\gamma}_{ab}\equiv\frac{d\phi}{dr}\gamma_{ab}\;,
\end{equation}
where
\begin{equation}\label{rFi}
\Phi^2=2\Omega_{D-2}\left[\lambda_1 r^{D-2}+
2(D-2)(D-3)\lambda_2 r^{D-4}\right]\;,
\end{equation}
and $q$ is arbitrary dimensionless parameter, the action
(\ref{gb2ef}) becomes
\begin{equation}\label{gb2ef2}
I_{\mbox{\scriptsize GB}}=\int d^2x\sqrt{-\tilde{\gamma}}
\left[\frac{1}{4}q\Phi_h\phi\tilde{\mathcal{R}}-V(\phi)\right].
\end{equation}
This action can be put in more familiar form if we make additional
conformal reparametrization:
\begin{equation}
\bar{\gamma}_{ab}\equiv e^{-\frac{2\phi}{q\Phi_h}}
\tilde{\gamma}_{ab}\;,
\end{equation}
Now (\ref{gb2ef2}) takes the form
\begin{equation}\label{gb2ef3}
I_{\mbox{\scriptsize GB}}=-\int d^2x\sqrt{-\bar{\gamma}}
\left[\frac{1}{2}(\bar{\nabla}\phi)^2-
\frac{1}{4}q\Phi_h\phi\bar{\mathcal{R}}+U(\phi)\right]\;,
\end{equation}
which is simmilar to the Liouville action. The difference is that
potential $U(\phi)$ is not purely exponential but is given with
\begin{displaymath} \!\!\!\!\!\!\!
U(\phi)=(D-2)(D-3)\Omega_{D-2}\left[\lambda_1r^{D-4}+
\lambda_2(D-4)(D-5)r^{D-6}\right]\frac{dr}{d\phi}
e^{\frac{2\phi}{q\Phi_h}}\;.
\end{displaymath}
Action (\ref{gb2ef3}) is of the same form as that obtained from pure
Einstein action. In \cite{Solod99} it was shown that if one imposes
condition that the metric $\bar{\gamma}_{ab}$ is \emph{nondynamical}
then the action (\ref{gb2ef3}) describes CFT \emph{near the
horizon}\footnote{Carlip showed that above condition is indeed
consistent boundary condition \cite{Carlip01}.}. We therefore fix
$\bar{\gamma}_{ab}$ near the horizon and take it to be metric of
static spherically symmetric black hole:
\begin{equation}\label{ds2w}
d\bar{s}_{(2)}^2\equiv\bar{\gamma}_{ab}dx^adx^b=
-f(w)dt^2+\frac{dw^2}{f(w)}\;,
\end{equation}
where near the horizon $f(w_h)=0$ we have
\begin{equation}\label{fwh}
f(w)=\frac{2}{\beta}(w-w_h)+O\left((w-w_h)^2\right)\;.
\end{equation}
We now make coordinate reparametrization $w\to z$
\begin{equation}\label{zw}
z=\int^w\frac{dw}{f(w)}=\frac{\beta}{2}\ln\frac{w-w_h}{f_0}+O(w-w_h)
\end{equation}
in which 2-dim metric has a simple form
\begin{equation}\label{ds2z}
d\bar{s}_{(2)}^2=f(z)\left(-dt^2+dz^2\right)\;,
\end{equation}
and the function $f$ behaves near the horizon ($z_h=-\infty$) as
\begin{equation}\label{fzh}
f(z)\approx f_0 e^{2z/\beta}\;,
\end{equation}
i.e., it \emph{exponentially} vanishes. It is easy to show that
equation of motion for $\phi$ which follows from Eqs. (\ref{gb2ef3}),
(\ref{ds2z}), (\ref{fzh}) is
\begin{equation}\label{eom}
\left(-\partial_t^2+\partial_z^2\right)\phi=\frac{1}{4}q\Phi_h
\bar{\mathcal{R}}f+fU'(\phi)\approx O\left(e^{2z/\beta}\right)\;,
\end{equation}
and that the ``flat'' trace of the energy-momentum tensor is
\begin{equation}\label{temt}
-T_{00}+T_{zz}=\frac{1}{4}q\Phi_h\left(-\partial_t^2+
\partial_z^2\right)\phi-fU(\phi)\approx O\left(e^{2z/\beta}\right)\;,
\end{equation}
which is exponentially vanishing near the horizon\footnote{Higher
derivative terms in (\ref{gb2ef}) make contribution to (\ref{temt})
proportional to $f(\nabla\phi)^2\approx o(\exp(2z/\beta))$.}. From
(\ref{eom}) and (\ref{temt}) follows that the theory of the scalar
field $\phi$ exponentially approaches CFT near the horizon.

Now, one can construct corresponding Virasoro algebra using standard
procedure. Using light-cone coordinates $z_\pm=t\pm z$ right-moving
component of energy--momentum tensor near the horizon is approximately
\begin{equation} \label{tempp}
T_{++}=(\partial_+\phi)^2 - \frac{1}{2}q\Phi_h\partial_+^2\phi + 
\frac{q\Phi_h}{2\beta}\partial_+\phi \;.
\end{equation}
It is important to notice that horizon condition (\ref{hoco}) implies
that $r$ and $\phi$ are (approximately) functions only of one
light-cone coordinate (we take it to be $z_+$), which means that only
one set of modes (left \emph{or} right) is contributing.

Virasoro generators are coefficients in the Fourier expansion of
$T_{++}$:
\begin{equation} \label{tfour}
T_n=\frac{\ell}{2\pi}\int_{-\ell/2}^{\ell/2}dz\,e^{i2\pi nz/\ell}
T_{++} \;,
\end{equation}
where we compactified $z$-coordinate on a circle of circumference
$\ell$.
Using canonical commutation relations it is easy to show that Poisson
brackets of $T_n$'s are given with
\begin{equation} \label{fvir}
i\{T_n,T_m\}_{\mbox{\scriptsize PB}}=(n-m)T_{n+m}
+\frac{\pi}{4}q^2\Phi_h^2\left(n^3+n
\left(\frac{\ell}{2\pi\beta}\right)^2\right)\delta_{n+m,0} \;.
\end{equation}
To obtain the algebra in quantum theory (at least in semiclassical
approximation) one replaces Poisson brackets with commutators using
$[\,,]=i\hbar\{\,,\}_{\mbox{\scriptsize PB}}$, and divide generators
by $\hbar$. From (\ref{fvir}) it follows that ``shifted'' generators
\begin{equation} \label{lntn}
L_n=\frac{T_n}{\hbar}+\frac{c}{24}\left(\left(
\frac{\ell}{2\pi\beta}\right)^2+1\right)\delta_{n,0} \;,
\end{equation}
where
\begin{equation} \label{cch}
c=3\pi q^2\frac{\Phi_h^2}{\hbar} \;,
\end{equation}
satisfy Virasoro algebra
\begin{equation} \label{vir}
[L_n,L_m]=(n-m)L_{n+m}+\frac{c}{12}\left(n^3-n\right)\delta_{n+m,0}
\end{equation}
with central charge $c$ given in (\ref{cch}).

Outstanding (and unique, as far as is known) property of the Virasoro
algebra is that in its representations a logarithm of the number of
states (i.e., entropy) with the eigenvalue of $L_0$ equal to $\Delta$ is
asymptoticaly given with Cardy formula (\ref{carfo}). 
If we assume that in our case $\Delta_g=0$ in semiclassical
approximation (more precisely, $\Delta_g\ll c/24$), one can see that
number of microstates (purely quantum quantity) is in leading
approximation completely determined by (semi)classical values of $c$
and $L_0$. Now it only remains to determine $\Delta$. In a classical
black hole solution we have
\begin{equation} \label{rsol}
r=w=w_h+(w-w_h)\approx r_h+f_0 e^{2z/\beta}\;,
\end{equation}
so from (\ref{rfi}) and (\ref{rFi}) follows that near the horizon
$\phi\approx\phi_h$.
Using this configuration in (\ref{tfour}) one obtains $T_0=0$, which
plugged in (\ref{lntn}) gives
\begin{equation} \label{delta}
\Delta=\frac{c}{24}
\left(\left(\frac{\ell}{2\pi\beta}\right)^2+1\right)\;.
\end{equation}
Finally, using (\ref{cch}) and (\ref{delta}) in Cardy formula
(\ref{carfo}) one obtains
\begin{equation} \label{sconf}
S_{\mbox{\scriptsize C}}=\frac{c}{12}\frac{\ell}{\beta}
=\frac{\pi}{4}q^2\frac{\ell}{\beta}\frac{\Phi_h^2}{\hbar}\;.
\end{equation}
Let us now compare (\ref{sconf}) with classical formula (\ref{sgb}),
which in present case is
\begin{equation} \label{sgb2}
S_{\mbox{\scriptsize GB}}=\frac{4\pi}{\hbar}\oint d^{D-2}x
\sqrt{\tilde{g}}\left(\lambda_1-2\lambda_2 R(\tilde{g}_{ij})\right)
\;,
\end{equation}
where $\tilde{g}_{ij}$ is induced metric on the horizon. In the
sphericaly symmetric case horizon is a $(D-2)$-dimensional sphere with
radius $r_h$ and $R(\tilde{g}_{ij})=-(D-2)(D-3)/r_h^2$, so
(\ref{sgb2}) becomes
\begin{equation} \label{sgb2s}
S_{\mbox{\scriptsize GB}}=\frac{4\pi}{\hbar}\Omega_{D-2}
\left[\lambda_1r_h^{D-2}+2(D-2)(D-3)\lambda_2r_h^{D-4}\right]
=2\pi\frac{\Phi_h^2}{\hbar}\;.
\end{equation}
Using this our expression (\ref{sconf}) can be written as
\begin{equation} \label{scsgb2}
S_{\mbox{\scriptsize C}}=
\frac{q^2}{8}\frac{\ell}{\beta}S_{\mbox{\scriptsize GB}}\;,
\end{equation}
so it gives correct result apart from dimensionless coeficient, which
can be determined in the same way as in pure Einstein case
\cite{Carlip01}. First, it is natural to set the compactification
period $\ell$ equal to period of Euclidean-rotated black
hole\footnote{We note that our functions depend only on variable
$z_+$, so the periodicity properties in time $t$ are identical to
those in $z$.}, i.e.,
\begin{equation} \label{lwick}
\ell=2\pi\beta \;.
\end{equation}
The relation between eigenvalue $\Delta$ of $L_0$ and $c$ then becomes
\begin{equation} \label{delc}
\Delta=\frac{c}{12} \;.
\end{equation}
We shall see in the next section that this relation holds for
general GB theory, i.e., for arbitrary values of coupling 
constants\footnote{For pure Einstein gravity this relation is
implicitely given in \cite{Carlip99}.}. One could be tempted to expect
this to be valid for larger class of black holes and interactions then
those treated so far.

To determine dimensionless parameter $q$ we note that our effective
theory given with (\ref{gb2ef3}) depends on effective parameters
$\Phi_h$ and $\beta$, and thus one expects that $q$ depends on
coupling constants only through dimensionless combinations of them.
Thus to determine $q$ one may consider $\lambda_2=0$ case and compare
expression for central charge (\ref{cch}) with that obtained in
\cite{Carlip99}, which is
\begin{equation} \label{cchcar}
c=\frac{3A_h}{2\pi\hbar G_D}\;,
\end{equation}
where $A_h=\Omega_{D-2}r_h^{D-2}$ is the area of horizon. One obtains
that
\begin{equation} \label{qcar}
q^2=\frac{4}{\pi} \;.
\end{equation}
One could also perform boundary analysis of Ref. \cite{Carlip99} for
GB gravity (see Appendix). This procedure gives
$\Delta=\Phi_h^2/\hbar$ which combined with (\ref{cch}) and
(\ref{delc}) gives (\ref{qcar}).

Using (\ref{lwick}) and (\ref{qcar}) one finally obtains desired
result
\begin{equation} \label{scsgb}
S_{\mbox{\scriptsize C}}=S_{\mbox{\scriptsize GB}} \;.
\end{equation}

\section{General Gauss--Bonnet gravity}

In $D>6$ Gauss--Bonnet action has additional terms and general action
was given in (\ref{igb}). Using spherical symmetry one obtains
effective two-dimensional action given now with
\begin{eqnarray} \label{gbef}
I_{\mbox{\scriptsize GB}}&=&\Omega_{D-2}\sum_{m=0}^{[D/2]}\lambda_m
\frac{(D-2)!}{(D-2m)!}\int d^2x\sqrt{-\gamma}\,r^{D-2m-2}
\left[1-(\nabla r)^2\right]^{m-2} \nonumber \\ &&\times \bigg\{
2m(m-1)r^2\left[(\nabla_a\nabla_br)^2-(\nabla^2r)^2\right] \nonumber\\
&&\quad+2m(D-2m)r\nabla^2r\left[1-(\nabla r)^2\right]
+m\mathcal{R}r^2\left[1-(\nabla r)^2\right] \nonumber \\
&&\quad\left. -(D-2m)(D-2m-1)\left[1-(\nabla r)^2\right]^2\right\}\;.
\end{eqnarray}
After partial integration\footnote{Notice that
\begin{eqnarray*}
2r^n\left[(\nabla_a\nabla_br)^2-(\nabla^2r)^2\right]&=&
3nr^{n-1}\nabla^2r(\nabla r)^2+n(n-1)r^{n-2}(\nabla r)^4 \\
&&+\mathcal{R}r^n(\nabla r)^2+\mbox{surface terms}\;.
\end{eqnarray*}} and implementation of horizon condition
$(\nabla r)^2\approx 0$, (\ref{gbef}) becomes near the horizon
approximately
\begin{eqnarray} \label{gbefh}
I_{\mbox{\scriptsize GB}}&=&-\Omega_{D-2}\sum_{m=0}^{[D/2]}\lambda_m
\frac{(D-2)!}{(D-2m-2)!}\int d^2x\sqrt{-\gamma}\,r^{D-2m-2}\nonumber\\
&&\times\left\{m(\nabla r)^2-\frac{m}{(D-2m)(D-2m-1)}\mathcal{R}r^2
+1\right\} \;.
\end{eqnarray}
If we define
\begin{equation} \label{rFiD}
\Phi^2\equiv 2\Omega_{D-2}\sum_{m=1}^{[D/2]}m\lambda_m
\frac{(D-2)!}{(D-2m)!}\,r^{D-2m} \;,
\end{equation}
and make a reparametrization (\ref{rfi}), the action (\ref{gbefh})
becomes
\begin{equation} \label{gbef2}
I_{\mbox{\scriptsize GB}}=\int d^2x\sqrt{-\tilde{\gamma}}
\left[\frac{1}{4}q\Phi_h\phi\tilde{\mathcal{R}}-V(\phi)\right]
\end{equation}
which is of the same form as (\ref{gb2ef2}) (the only difference is
the exact form of the potential which is unimportant in this
calculation). Now one can repeat the analysis from (\ref{gb2ef2}) to
(\ref{sconf}) in previous section without a change and obtain for the
entropy the expression (\ref{sconf}), where $\Phi_h$ is now given by
(\ref{rFiD}) evaluated at a horizon. It only remained to show that
also in the general case the entropy (\ref{sgb}) and $\Phi_h$ are
related as in (\ref{sgb2s}). For spherically symmetric metric
(\ref{gsph}) where horizon is a $(D-2)$-dimensional sphere with radius
$r_h$ one can show that (\ref{sgb}) can be written as\footnote{In fact
it is obvious that the last term in (\ref{gbefh}) is minus $m$-th
Gauss--Bonnet density (\ref{lgbm}) for the $(D-2)$-dimensional sphere
with radius $r$, i.e.,
\begin{displaymath}
\mathcal{L}_m=\frac{(D-2)!}{(D-2m-2)!}\Omega_{D-2}\,r^{D-2m-2} \;.
\end{displaymath}}
\begin{equation} \label{sgbss}
S_{\mbox{\scriptsize GB}}=\frac{4\pi}{\hbar}\Omega_{D-2}
\sum_{m=1}^{[D/2]}m\lambda_m\frac{(D-2)!}{(D-2m)!}\,r^{D-2m}
=2\pi\frac{\Phi_h^2}{\hbar}
\end{equation}
the same as in (\ref{sgb2s}). Finally, using the same arguments as in
previous section one obtains
$S_{\mbox{\scriptsize C}}=S_{\mbox{\scriptsize GB}}$.

\section{Conclusion}

In this Letter we have calculated entropy of $D$-dimensional
spherically symetric black holes in Gauss--Bonnet gravity. The method
used asymptotic conformal symmetry of the effective two-dimensional
action near the horizon \cite{Solod99}. This makes it possible to find
via Cardy formula number of microstates. The obtained relation for the
entropy coincides with the macroscopic formula \cite{JacMye93}.

It would be desirable to investigate if this result pursues also in
other interactions. It would be also of interest to treat a
more general class of stationary black holes. Such questions maybe
also addressed by Carlip methods \cite{Carlip99}. This could also help
to understand better the relation of two methods and the question of
their eventual equivalence (some progress in this direction was
recently done in \cite{Carlip02}). In fact, some of these questions
will be addressed in a separate publication.

\section*{Acknowledgements}
Two of us (S.\ P. and P.\ P.) would like to acknowledge the kind
hospitality of SISSA High Energy Division where part of the work
was done. We would like also to acknowledge the financial support
under the contract of Italo-Croatian collaboration and the contract
No. 119222 of Ministery of Science and Technology of Republic of
Croatia.

\section*{Appendix}

Eigenvalue $\Delta$ of $L_0$ can be calculated by boundary analysis
previously applied for Einstein gravity by Carlip\footnote{We use here
the notation of Ref. \cite{Carlip99} where possible.} \cite{Carlip99}.
It is a contribution to the boundary term of the Hamiltonian
\begin{displaymath}
H[\xi]=\int_{\mathcal{H}}Q[\xi]+\ldots \;.
\end{displaymath}
Here $\mathcal{H}$ denotes the $(n-2)$-dimensional intersection of the
Cauchy surface with the horizon, and the $(n-2)$-form $Q$ is equal to
\begin{displaymath}
Q_{a_3\cdots a_n}[\xi]=-\frac{\partial
\mathcal{L}_{\mbox{\scriptsize GB}}}{\partial R_{abcd}}
\eta_{ab}\nabla_{[c}\xi_{d]}\,\hat{\epsilon}_{a_3\cdots a_n} \;,
\end{displaymath}
$\eta_{ab}$ is the binormal to the $\mathcal{H}$, and $\xi^a$ is the
vector field to which corresponds generator of diffeomorphisms
$H[\xi]$. Boundary and integrability conditions are fixing
deformations to lie in ``$r$-$t$'' plane and $\xi^a=K\rho^a+T\chi^a$,
where $\chi^a$ is approximately Killing near the horizon (determined by 
$\chi^2=0$), and $\rho_a=-\nabla_a\chi^2/2\kappa$. Scalars $K$, $T$ are
connected by $K=\chi^2\chi^a\nabla_a T/\kappa\rho^2\equiv
\dot{T}\chi^2/\kappa\rho^2$. Then one can calculate
\begin{equation}\label{app1}
J[\xi]\equiv\int_{\mathcal{H}}Q[\xi]
=\int_{\mathcal{H}}\left(\lambda_1-2\lambda_2\,^{(n-2)}R\right)
\left(2\kappa T-\frac{\ddot{T}}{\kappa}\right)
\hat{\epsilon}_{a_3\cdots a_n} \;.
\end{equation}
One can show analogously to \cite{Carlip99} that Fourier components
$T_m$ of $T$ lead to generators $J[T_m]$ whose Dirac
brackets satisfy again Virasoro algebra. Eq. (\ref{app1}) then gives
us
\begin{displaymath}
\hbar\Delta=J[T_0]=\int_{\mathcal{H}}
\left(2\lambda_1-4\lambda_2\,^{(n-2)}R\right)
\hat{\epsilon}_{a_3\cdots a_n}=\Phi_h^2 \;.
\end{displaymath}
Comparison with (\ref{delc}) gives (\ref{qcar}). All details will be
given in a separate publication \cite{CvPaPPip}.


\begin{thebibliography}{99}

\bibitem{BekHaw}
J. D. Bekenstein, {\em Lett. Nuovo. Cim.\/} {\bf 4} (1972) 737;
  {\em Phys. Rev. D\/} {\bf 7} (1973) 2333;
  {\em Phys. Rev. D\/} {\bf 9} (1974) 3292.
S. W. Hawking, {\em Nature\/} {\bf 248} (1974) 30;
  {\em Commun. Math. Phys.\/} {\bf 43} (1975) 199.
\bibitem{BroHen86}
J. D. Brown and M. Henneaux, {\em Commun. Math. Phys.\/} {\bf 104}
  (1986) 207.
\bibitem{BanStro}
M. Banados, C. Teitelboim and J. Zanelli, {\em Phys. Rev. Lett.\/}
  {\bf 69} (1992) 1849;
A. Strominger, {\em JHEP\/} {\bf 9802} (1998) 009.
\bibitem{Solod99}
S. N. Solodukhin, {\em Phys. Lett.\/} {\bf B454} (1999) 213.
\bibitem{Cardy}
J. A. Cardy, {\em Nucl. Phys.\/} {\bf B270} (1986) 186;
H. W. J. Bl\"{o}te, J. A. Cardy and M. P. Nightingale, 
  {\em Phys. Rev. Lett.\/} {\bf 56} (1986) 742.
\bibitem{Carlip98}
S. Carlip, {\em Phys. Rev. Lett.\/} {\bf 82} (1999) 2828;
V. O. Soloviev, {\em Phys. Rev. D\/} {\bf 61} (1999) 027502.
\bibitem{Carlip99}
S. Carlip, {\em Class. Quant. Grav.\/} {\bf 16} (1999) 3327.
\bibitem{Carlip02}
S. Carlip, {\em Phys. Rev. Lett.\/} {\bf 88} (2002) 241301.
\bibitem{Park01}
M.-I. Park, hep-th/0111224.
\bibitem{Silva02}
S. Silva, hep-th/0204179.
\bibitem{Lovelock}
D. Lovelock, {\em J. Math. Phys.\/} {\bf 12} (1971) 498;
  {\em J. Math. Phys.\/} {\bf 13} (1972) 874.
\bibitem{JacMye93}
T. Jacobson and R. C. Myers, {\em Phys. Rev. Lett.\/} {\bf 70} (1993) 
  3684.
\bibitem{genent}
R. M. Wald, {\em Phys. Rev. D\/} {\bf 48} (1993) 3427;
V. Iyer and R. M. Wald, {\em Phys. Rev. D\/} {\bf 50} (1994) 846;
T. Jacobson, G. Kang and R. C. Myers, {\em Phys. Rev. D\/} {\bf 49}
  (1994) 6587.
\bibitem{noghost}
B. Zwiebach, {\em Phys. Lett.\/} {\bf B156} (1985) 315;
B. Zumino, {\em Phys. Rept.\/} {\bf 137} (1986) 109.
\bibitem{ChNeWe02}
Y.M. Cho, I. P. Neupane and P.S. Wesson, {\em Nucl. Phys.\/}
  {\bf B621} (2002) 388.
\bibitem{Myers87}
R. C. Myers, {\em Phys. Rev. D\/} {\bf 36} (1987) 392.
\bibitem{MyeSim88}
R. C. Myers and J. Z. Simon, {\em Phys. Rev. D\/} {\bf 38} (1988)
  2434;
R. G. Cai, {\em Phys. Rev. D\/} {\bf 65} (2002) 084014;
R. G. Cai and K. S. Soh, {\em Phys. Rev. D\/} {\bf 59} (1999) 044013.
\bibitem{CvNoOd02}
M. Cveti\v{c}, S. Nojiri and S. D. Odintsov, {\em Nucl. Phys.\/}
  {\bf B628} (2002) 295.
\bibitem{Russo95}
J. G. Russo, {\em Phys. Lett.\/} {\bf B359} (1995) 69.
\bibitem{Carlip01}
S. Carlip, {\em Phys. Lett.\/} {\bf B508} (2001) 168.
\bibitem{CvPaPPip}
M. Cvitan, S. Pallua and P. Prester, in preparation.

\end{thebibliography}
\end{document}